\documentclass[%
preprint,
amsmath,amssymb,
]{revtex4-1}

\usepackage{graphicx}% Include figure files
\usepackage{dcolumn}% Align table columns on decimal point
\usepackage{bm}% bold math
\makeatletter
\def\maketitle{ %
\say\@authors
\@author@finish
\title@column\titleblock@produce
\suppressfloats[t]%
\let\and\relax
\let\affiliation\@gobble@opt@one
\let\address\affiliation
\let\author\@gobble
\@author@init
\let\@authors\@empty
\let\@authors@curr\@empty
\let\@affil@list\@empty
\let\keywords\@gobble
\let\@keywords\@empty
\let\email\@gobble
\let\@address\@empty
\let\maketitle\relax
\let\thanks\@gobble
\titlepage@sw{ %
\clearpage
 }{}%
 }%
 \makeatother

\begin{document}
%\renewcommand{\figurename}{Figure}
%\preprint{APS/123-QED}

\title{Spin-scattering asymmetry at half-metallic ferromagnet/ferromagnet interface}% Force line breaks with \\

\author{Y. Fujita,$^{1}$$\footnote{E-mail: FUJITA.Yuichi@nims.go.jp}$  Y. Miura,$^{2}$ T. Sasaki,$^{2}$ T. Nakatani,$^{2}$ K. Hono,$^{2}$ and Y. Sakuraba$^{2}$$\footnote{E-mail: SAKURABA.Yuya@nims.go.jp}$}
\affiliation{
$^{1}$International Center for Young Scientists, National Institute for Materials Science, 1-2-1 Sengen, Tsukuba, Ibaraki 305-0047, Japan. 
}
\affiliation{
$^{2}$Research Center for Magnetic and Spintronic Materials, National Institute for Materials Science, 1-2-1 Sengen, Tsukuba, Ibaraki 305-0047, Japan. 
}

\date{\today}% It is always \today, today,

\begin{abstract}
We study spin-scattering asymmetry at the interface of two ferromagnets (FMs) based on a half-metallic Co$_{2}$Fe$_{0.4}$Mn$_{0.6}$Si (CFMS)/CoFe interface. First-principles ballistic transport calculations based on Landauer formula for (001)-CoFe/CFMS/CoFe indicate strong spin-dependent conductance at the CFMS/CoFe interface, suggesting a large interface spin-scattering asymmetry coefficient ($\gamma$).  Fully epitaxial current-perpendicular-to-plane giant magnetoresistance (CPP-GMR) pseudo-spin-valve (PSV) devices involving CoFe/CFMS/Ag/CFMS/CoFe structures exhibit an enhancement in MR output owing to the formation of the CFMS/CoFe interface at room temperature (RT). This is well reproduced qualitatively by a simulation based on a generalized two-current series-resistor model with considering the presence of $\gamma$ at the CFMS/CoFe interface, half-metallicity of CFMS, and combinations of terminated atoms at the interfaces in the CPP-GMR PSV structure. We show direct evidence for a large $\gamma$ at a half-metallic FM/FM interface and its impact on CPP-GMR effect even at RT. 
\end{abstract}

\maketitle

%%%%%%%%%%%%%%%%%%%%
%\section{Introduction}
%%%%%%%%%%%%%%%%%%%%

Spin-dependent transport of conduction electrons at metal/metal interfaces, which is involved in the giant magnetoresistance (GMR) effect \cite{Baibich_PRL, Pratt_PRL, Camley_PRL, Parkin_PRL, Hsu_PRL, Borchers_PRL, Bozec_PRL}, has been a longstanding subject of great interest. 
The interface spin-scattering asymmetry coefficient ($\gamma$) defined as $\gamma = \frac{R^{\downarrow}A - R^{\uparrow}A}{R^{\downarrow}A + R^{\uparrow}A}$, where $R^{\uparrow}A$ and $R^{\downarrow}A$ are the resistance area product ($RA$) for majority- and minority-spin channels, respectively, at ferromagnet (FM)/nonmagnet (NM) interfaces has been evaluated for various combinations of FM and NM metals \cite{Bass_JPhys_review, Bass_JMMM_review} through the experimental analysis of resistance change-area product ($\Delta RA$) observed in the current-perpendicular-to-plane GMR (CPP-GMR) devices in terms of the two-current series-resistor (2CSR) model \cite{Lee_JMMM, VF_PRB}.  
In addition, the first-principles theories revealed that $\gamma$ at FM/NM interfaces originates from the spin-dependent matching of interfacial band dispersions \cite{Schep_PRL, Bass_JMMM_review, Butler_PRB, Stiles_JAP, Schep_PRB}.

As in the case of FM/NM interfaces, $\gamma$ can be yielded even at FM/FM interfaces because they have spin-dependent interfacial band matching. 
However, it is difficult to evaluate $\gamma$ at FM/FM interfaces because of the strong magnetic exchange coupling at FM/FM interfaces. 
Nguyen {\it et al.} experimentally observed spin-dependent scattering at the Co/Ni interface at 4.2 K via the analysis of $\Delta RA$ observed in the CPP-GMR devices with several [Co/Ni]$_{n}$ superlattices and the theoretical calculations \cite{CoNi_PRB}. 
Interestingly, the experimental (theoretical) value of $\gamma$ at the Co/Ni interface was estimated to be 0.94 (0.97), which was considerably larger than that at FM/NM interfaces with all the combinations of FM and NM that have been verified so far \cite{Bass_JMMM_review}. 
Based on $\gamma$ at the Co/Ni interface, one can expect that there are FM/FM interfaces that give rise to considerably large $\gamma$ at room temperature (RT). However, there is no report on the spin-dependent scattering at FM/FM interfaces at RT.

From the viewpoint of an electronic structure, half-metallic FM (HMF)/FM interfaces offer strong spin-dependent scattering because HMFs have a semiconducting gap only in either a majority- or a minority-spin band. 
The Co-based full Heusler alloys, such as Co$_{2}$Fe$_{x}$Mn$_{1-x}$Si and Co$_{2}$FeGa$_{x}$Ge$_{1-x}$ ($0\leq x \leq 1$), which have theoretically been predicted to exhibit half metallicity \cite{Balke_PRB, Ozdogan_JAP, Varaprasad_Acta}, are the most widely explored HMFs for CPP-GMR devices \cite{Furubayashi_APL, Iwase_APL, Nakatani_APL, Sakuraba_PRB, Sakuraba_APL, Li_APL, Jung_APL, Li_APL2, Kubota_JPD, Inoue_APL, Bjoern_PRBL} and $\gamma$ at the Co-based Heusler alloy/NM interfaces has been well-documented \cite{Sakuraba_PRB, Miura_PRB, Jung_APL, Bjoern_PRBL, Kubota_JPD}. 
The experiments with the CPP-GMR devices and the theories based on the first-principles calculations for interfacial band matching and overlapping of Fermi surfaces have confirmed that $\gamma$ at Co$_{2}$Fe$_{x}$Mn$_{1-x}$Si/NM ($x =$ 0 and 0.4) interfaces strongly depends on the material of the NM \cite{Sakuraba_PRB, Miura_PRB, Kubota_JPD}. 
Furthermore, the enhancement of $\gamma$ at the Co$_{2}$FeGa$_{0.5}$Ge$_{0.5}$ (CFGG)/Ag interface was demonstrated by inserting an ultrathin ($<$ 1 nm) NiAl layer or Ni layer into the CFGG/Ag interface; the theoretical calculation suggests that the enhancement is attributed to the considerable improvement in interfacial band matching \cite{Jung_APL, Bjoern_PRBL}.
Thus, the Co-based Heusler alloys are promising for exploring spin-dependent scattering even at HMF/FM interfaces. 

In this Letter, we present evidence for spin-scattering asymmetry at the half-metallic Co$_{2}$Fe$_{0.4}$Mn$_{0.6}$Si (CFMS)/CoFe interface based on the first-principles ballistic transport calculations for (001)-CoFe/CFMS/CoFe and MR measurements of fully epitaxial CPP-GMR pseudo spin valves (PSVs) with CFMS($t$ nm)/CoFe($7 - t$ nm) ($0\leq t \leq 7$) layers. We show direct evidence for the presence of a large $\gamma$ at the CFMS/CoFe interface and its impact on the GMR effect at RT by verifying the $t$ dependence of $\Delta RA$.

\begin{figure*}
\begin{center}
\includegraphics[width=15cm]{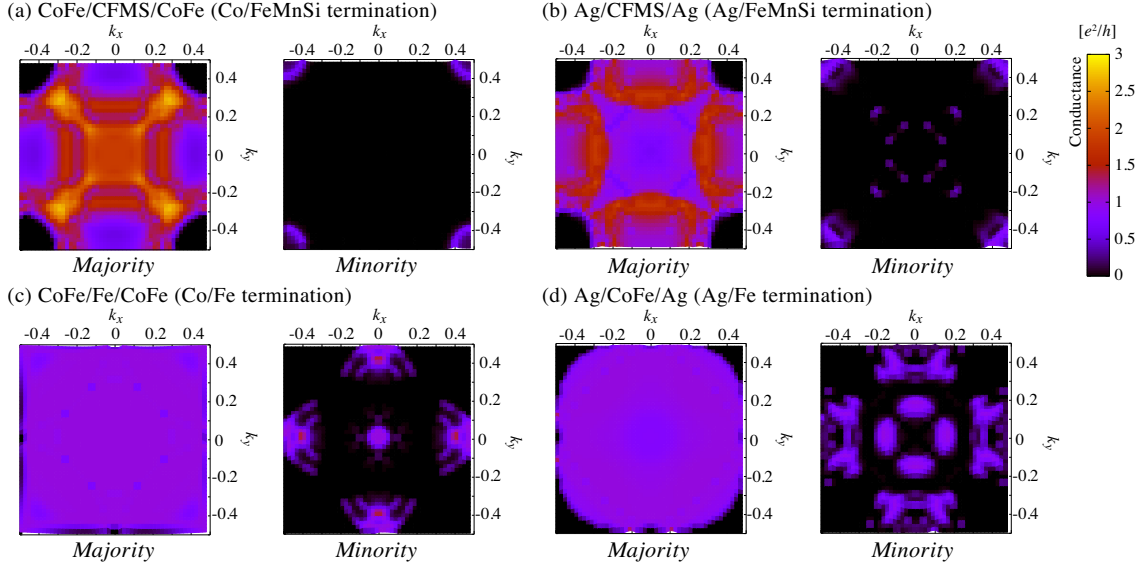}
\caption{Majority-spin (left) and minority-spin (right) conductance at the Fermi energy calculated for (a) (001)-CoFe/CFMS/CoFe with Co-terminated CoFe and FeMnSi-terminated CFMS layers, (b) (001)-Ag/CFMS/Ag with an FeMnSi-terminated CFMS layer, (c) (001)-CoFe/Fe/CoFe with Co-terminated CoFe layers, and (d) (001)-Ag/CoFe/Ag with an Fe-terminated CoFe layer as functions of $\bm k_{\parallel} = (k_{x}, k_{y})$. For the calculations of (a) and (c), the magnetization configurations are set to be parallel.}
\end{center}
\end{figure*}

First of all, we performed first-principles ballistic transport calculations based on the Landauer formula for (001)-CoFe/CFMS/CoFe to explore the existence of spin-dependent scattering at the CFMS/CoFe interface.
We used QUANTUM-ESPRESSO code for the electronic structure and transport calculations \cite{Smogunov_PRB, Giannozzi_JPhys} with the generalized gradient approximation for the exchange and correlation terms \cite{Perdew_PRL}.
The details of the calculation method are explained in Ref. \citenum{Miura_PRB}. 
The open quantum system comprises a tetragonal supercell containing 13 atomic layers of CFMS with Co and FeMnSi termination and 7 atomic layers of CoFe with Co and Fe termination. 
The interface distances of CoFe/CMFS and Ag/CMFS junctions were determined by structure relaxations in the QUANTUM-ESPRESSO code. 
The convergence criteria of the force are less than 10$^{-5}$ Rydberg/Bohr for each atomic position. 
We performed the optimizations by changing the initial interface distance of super-cells, and determined the optimal interface distance for each termination. 
In the ballistic transport of magnetic junctions, conduction electrons scattering occurs due to the potential energy near the interfaces. 
As references, we performed the same calculations for (001)-Ag/CFMS/Ag, (001)-Ag/CoFe/Ag, and (001)-CoFe/Fe/CoFe with all combinations of terminated atoms. 
The magnetization configurations for CoFe/CFMS/CoFe and CoFe/Fe/CoFe were set to be parallel.
The number of in-plane $k$ points was considered 50 $\times$ 50 in the two-dimensional Brillouin zone (BZ) for ballistic conductance calculations.
Figures 1(a)-1(d) show the in-plane wave vector ($\bm k_{\parallel}$) dependencies of the majority-spin (left) and minority-spin (right) conductance at the Fermi energy normalized by $e^{2}/h$ averaged over the two-dimensional (2D) BZ for CoFe/CFMS/CoFe, Ag/CFMS/Ag, CoFe/Fe/CoFe, and Ag/CoFe/Ag, respectively, for each combination of terminated atoms as representatives. 
Figures 1(a) and 1(b) indicate that the highly conductive channels are distributed in almost the entire region of $\bm k_{\parallel} = (k_{x}, k_{y})$ in 2D BZ for the majority-spin channels for both CoFe/CFMS/CoFe and Ag/CFMS/Ag. 
In contrast to the majority-spin paths, the minority-spin paths have conductive channels with considerably small conductivity only around $\bm k_{\parallel} = (\pm 0.5, \pm 0.5)$. This indicates that $RA$ for the minority-spin channel is much larger than that for the majority-spin channel and suggests a large $\gamma$. 
Although Figs. 1(c) and 1(d) show a similar difference between majority- and minority-spin $\bm k_{\parallel}$ dependencies of conductance for CoFe/Fe/CoFe and Ag/CoFe/Ag unlike the cases for CoFe/CFMS/CoFe and Ag/CFMS/Ag, the conductive channels distinctly appear in the $\bm k_{\parallel}$ dependence of the minority-spin conductance around $\bm k_{\parallel} = (0, 0), (\pm0.5, 0)$ and $(0, \pm0.5)$ for CoFe/Fe/CoFe, and for some areas in $\bm k_{\parallel}$ for Ag/CoFe/Ag. 
The qualitative difference in the minority-spin conductance between structures with and without CFMS is attributed to the half metallicity of CFMS.
Note that qualitatively the same results as Fig. 1 were obtained for all the combinations of terminated atoms (not shown here).
$R^{\uparrow}A$ ($R^{\downarrow}A$) values for CoFe/CFMS/CoFe, Ag/CFMS/Ag, CoFe/Fe/CoFe, and Ag/CoFe/Ag averaged in each atomic termination can be calculated to be 3.18, 4.53, 2.30, and 2.62 m$\Omega$ $\mu$m$^{2}$ (243, 185, 18.4, and 21.5 m$\Omega$ $\mu$m$^{2}$), respectively, from the results of spin-dependent conductance; $\gamma$ at CoFe/CFMS, Ag/CFMS, CoFe/Fe, and Ag/CoFe interfaces can be estimated to be 0.97, 0.95, 0.78, and 0.78, respectively. 
The presence of large $\gamma$ is suggested even at CoFe/CFMS and the interfaces with half-metallic CFMS should yield much larger $\gamma$ than those without it.
The difference of $R^{\uparrow}A$ between the systems with CFMS, namely, CoFe/CFMS/CoFe and Ag/CFMS/Ag, is originated from the higher conductance around $\bm k_{\parallel} = (0, 0)$ and $(\pm0.3, \pm0.3)$ in CoFe/CFMS/CoFe than that in Ag/CFMS/Ag.
This should be originated from the larger overlapping of Fermi surfaces on the $\bm k_{\parallel}$ between CFMS and CoFe than that between CFMS and Ag as described in the supplemental material \cite{Supplemental}.
Further, the better band matching due to the relatively small interfacial distance of CoFe/CFMS(001) ($\sim$1.45 \AA) compared to that for Ag/CFMS(001) ($\sim$1.85 \AA), which is caused by the strong bonding between Co-Fe atoms, should be another reason for the smaller $R^{\uparrow}A$ value for CoFe/CFMS/CoFe than for Ag/CFMS/Ag.
%On the other hand, since $R^{\downarrow}A$ values for real metallic multilayered systems including HMF/metal interfaces can easily be fluctuated by uncontrollable origins such as atomic disorder, thermal fluctuation of magnetic moments, spin-orbit coupling, and so on, there are doubts about quantitative reliability of the above values of $R^{\downarrow}A$, and hence the estimated $\gamma$ values may have quantitative uncertainty.
%The relatively small interfacial distance of CoFe/CFMS(001) ($\sim$1.45 \AA) caused by the strong bonding between Co-Fe atoms is a possible reason for better band matching compared to that for Ag/CFMS(001) ($\sim$1.85 \AA).
Although $R^{\downarrow}A$ is significantly larger than $R^{\uparrow}A$ for both CoFe/CFMS/CoFe and Ag/CFMS/Ag, $R^{\downarrow}A$ at real systems with CFMS can easily be reduced by degradation of its half-metallic nature caused by atomic disorder, interface dislocation, thermal fluctuation of magnetic moments, spin-orbit coupling, and so on.
Therefore, it is suggested that experimentally obtainable $\gamma$ at the CoFe/CFMS interface is expected to be larger than that at the Ag/CFMS interface due to the smaller $R^{\uparrow}A$ for CoFe/CFMS/CoFe than that for Ag/CFMS/Ag.

\begin{figure}
\begin{center}
\includegraphics[width=8.5cm]{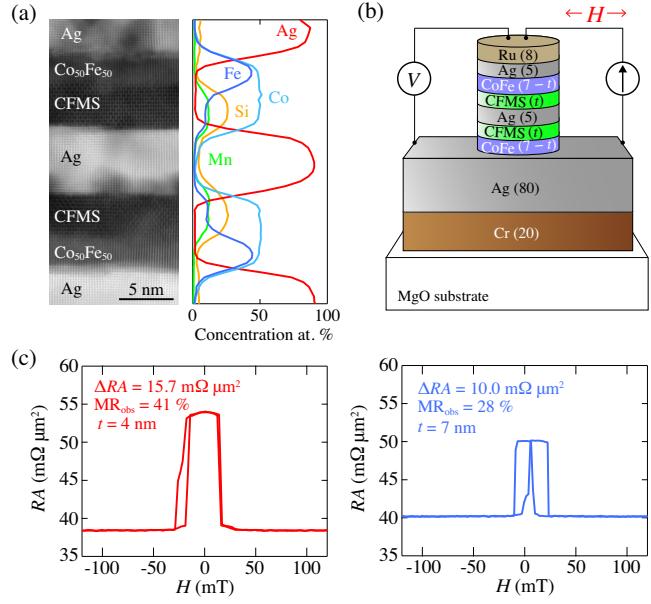}
\caption{(a) Cross-sectional HAADF-STEM image (left) and EDS line profiles (right) of the PSV ($t =$ 4 nm). (b) Schematic of the CPP-GMR PSV with CoFe($7 - t$ nm)/CFMS($t$ nm) ($0\leq t \leq 7$) structures. (c) Representative MR curves for devices with $t =$ 4 nm (left) and $t =$ 7 nm (right).}
\end{center}
\end{figure}

For examining the impact of $\gamma$ at the CFMS/CoFe interface ($\gamma_{\rm CFMS/CoFe}$) on the CPP-GMR effect, we fabricated fully epitaxial CoFe/CFMS/Ag/CFMS/CoFe structured CPP-GMR PSVs, wherein $\gamma_{\rm CFMS/CoFe}$ is expected to contribute to the GMR effect when the thickness of the CFMS layer is shorter than its spin diffusion length. 
We designed the PSV with the entire structure of MgO(001) single-crystalline substrate/Cr(20 nm)/Ag(80 nm)/Co$_{50}$Fe$_{50}$($7 - t$ nm)/CFMS($t$ nm)/Ag(5 nm)/CFMS($t$ nm)/Co$_{50}$Fe$_{50}$($7 - t$ nm)/Ag(5 nm)/Ru(8 nm) with various $t$ ($t =$ 0, 0.75, 1.5, 3, 4, 5, and 7 nm).
The PSVs with $t =$ 0 nm (CoFe/Ag/CoFe) and $t =$ 7 nm (CFMS/Ag/CFMS) have no CFMS/CoFe interfaces.
The samples were fabricated by processes with DC and RF sputtering at RT and $in$-$situ$ postdeposition annealing in an ultrahigh vacuum sputtering system with base pressures of $\sim$10$^{-6}$ Pa.
We confirmed the (001)-oriented fully epitaxial growth of the PSVs by x-ray diffraction (XRD) and the $L$2$_{1}$ ordered phase in the CFMS layers by nanobeam electron diffraction (NED).
From the cross-sectional high-angle annular dark-field scanning transmission electron microscope (HAADF-STEM) image and the energy dispersive x-ray spectroscopy (EDS) line concentration profile in Fig. 2(a), we can confirm that the CoFe/CFMS/Ag/CFMS/CoFe PSV structure is fabricated as designed with no evident atomic interdiffusion and has atomically flat and smooth interfaces.
The detailed procedure of the growth and results of XRD and NED and the detailed EDS elemental maps are summarized in the supplemental material \cite{Supplemental}. 
The atomic-resolution HAADF-STEM image shows that both Co and [(Fe, Mn), Si] atomic layers randomly exist as termination layers at CFMS/Ag interfaces (not shown here).
Further, we confirmed alternatively stacked Co and [(Fe, Mn), Si] atomic layers and $B$2-ordered CoFe near the CFMS/CoFe interfaces; hence, the CFMS/CoFe interface comprises both Co-terminated CFMS/Fe-terminated CoFe and [(Fe, Mn), Si]-terminated CFMS/Co-terminated CoFe.
Conventional electron-beam lithography and Ar$^{+}$ milling were used to fabricate circle- and ellipse-shaped pillar-type CPP-GMR devices as shown in Fig. 2(b). 
The designed size of the pillars ranged from 0.003 $\mu$m$^{2}$ to 0.03 $\mu$m$^{2}$ and their actual sizes were measured by analyzing the scanning electron microscope images of the pillars as the value of $A$. 
MR measurements were conducted by a DC four-probe method with a constant current of 1 mA at RT. 
The observed MR ratio (MR$_{\rm obs}$) and $\Delta$$RA$ are defined as $\frac {R_{\rm ap} - R_{\rm p}}{R_{\rm p}}$ and $(R_{\rm ap} - R_{\rm p})A$, respectively, where $R_{\rm p}$ and $R_{\rm ap}$ denote the resistance for parallel and antiparallel magnetization states between the top and the bottom ferromagnetic layers, respectively.
Because the parasitic lead resistance ($R_{\rm para}$) overlaps with device resistance and reduces the MR ratio, the intrinsic MR ratio (MR$_{\rm int}$) is defined as $\frac {\Delta R}{R_{\rm p} - R_{\rm para}}$.
$R_{\rm p}A$ and averaged MR$_{\rm int}$ can be estimated from the slope of an $R_{\rm p}$ vs. $1/A$ plot and a $\Delta$$R$ vs. $R_{\rm p}$ plot, respectively. 
The averaged $\Delta$$RA$ can be evaluated from MR$_{\rm int}$ $\times$ $R_{\rm p}A$ or the slope of a $\Delta$$R$ vs. $1/A$ plot.
Figure 2(c) shows the representative MR curves with clear plateaus for antiparallel magnetization states observed in the devices with $t =$ 4 nm (left) and $t =$ 7 nm (right), and with a designed pillar size of 0.015 $\mu$m$^{2}$. 
MR$_{\rm obs}$ and $\Delta$$RA$ for $t =$ 4 nm (41 $\%$ and 15.7 m$\Omega$ $\mu$m$^{2}$) are clearly larger than those for $t =$ 7 nm (28 $\%$ and 10.0 m$\Omega$ $\mu$m$^{2}$). 
Thus, we can tentatively state that introducing the CFMS/CoFe structure improves the MR output. 
To obtain averaged MR$_{\rm int}$ and $\Delta$$RA$, we performed MR measurements for 50 or more pillars for each $t$ with various $A$ as shown in Supplemental Fig. S3 in the supplemental material \cite{Supplemental}. 

\begin{figure}[t]
\begin{center}
\includegraphics[width=8.5cm]{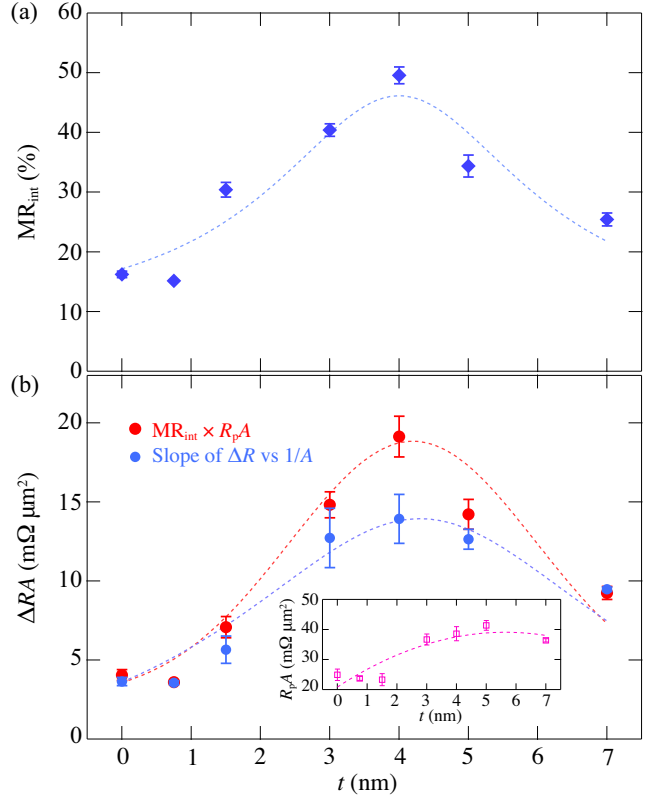}
\caption{$t$ dependence of (a) MR$_{\rm int}$ and (b) $\Delta$$RA$ estimated from MR$_{\rm int}$ $\times$ $R_{\rm p}A$ (red) and the slope of $\Delta$$R$ vs. $1/A$ plots (blue) for the PSVs with CFMS($t$ nm)/CoFe($7 - t$ nm) ($0\leq t \leq 7$) layers. The inset shows measured $R_{\rm p}A$ for each $t$. The dashed lines are merely guide to eye.}
\end{center}
\end{figure}

Figure 3(a) shows the averaged MR$_{\rm int}$ as functions of $t$.
A peak, where MR$_{\rm int}$ reached 50 $\%$, clearly appeared at $t =$ 4 nm.
A similar $t$ dependence of MR$_{\rm int}$ was observed for $\Delta$$RA$, as shown in Fig. 3(b). 
Although there is an apparent difference between $\Delta$$RA$ values estimated from MR$_{\rm int}$ $\times$ $R_{\rm p}A$ and $\Delta$$R$ vs. $1/A$ plots, clear peaks of $\Delta$$RA$ are observed irrespective of the estimation methods.
$R_{\rm p}A$ as a function of $t$ used to estimate $\Delta$$RA$ is shown in the inset of Fig. 3(b). 
In the supplemental material \cite{Supplemental}, we show plots to estimate MR$_{\rm int}$, $R_{\rm p}A$, and $\Delta$$RA$ and confirm that the quantitative difference in $\Delta$$RA$ is attributed to unavoidable errors in evaluating $A$ based on the estimation methods. 
The $\Delta$$RA$ values for $t =$ 0 nm, $t =$ 4 nm, and $t =$ 7 nm estimated from MR$_{\rm int}$ $\times$ $R_{\rm p}A$ ($\Delta$$R$ vs. $1/A$ plots) were 4.05 m$\Omega$ $\mu$m$^{2}$ (3.65 m$\Omega$ $\mu$m$^{2}$), 19.1 m$\Omega$ $\mu$m$^{2}$ (14.0 m$\Omega$ $\mu$m$^{2}$), and 9.25 m$\Omega$ $\mu$m$^{2}$ (9.48 m$\Omega$ $\mu$m$^{2}$), respectively.
According to the 2CSR model for the simple FM/NM/FM structure \cite{VF_PRB}, the magnitude of $\Delta$$RA$ can be increased by enhancing the bulk spin-scattering asymmetry coefficient ($\beta$) of FM defined as $\beta = \frac{\rho^{\downarrow} - \rho^{\uparrow}}{\rho^{\downarrow} + \rho^{\uparrow}}$, where $\rho^{\uparrow}$ and $\rho^{\downarrow}$ are the resistivities for majority- and minority-spin channels, respectively, and/or $\gamma$ at FM/NM interfaces.
Although the larger $\Delta$$RA$ for $t =$ 7 nm compared to that for $t =$ 0 nm is attributed to the enhancement of the $\beta$ of FM and $\gamma$ at the FM/NM interface by changing the FM from CoFe to half-metallic CFMS, the entire $t$-dependence of $\Delta$$RA$ in Fig. 3(b) cannot be explained only by considering the contribution of $\beta$ of CFMS layers ($\beta_{\rm CFMS}$) and $\gamma$ at the CFMS/Ag interface ($\gamma_{\rm CFMS/Ag}$).
The enhancement of $\Delta$$RA$ by employing CFMS/CoFe and the peak in its $t$-dependence can be attributed to $\gamma_{\rm CFMS/CoFe}$.

\begin{figure}
\begin{center}
\includegraphics[width=8.5cm]{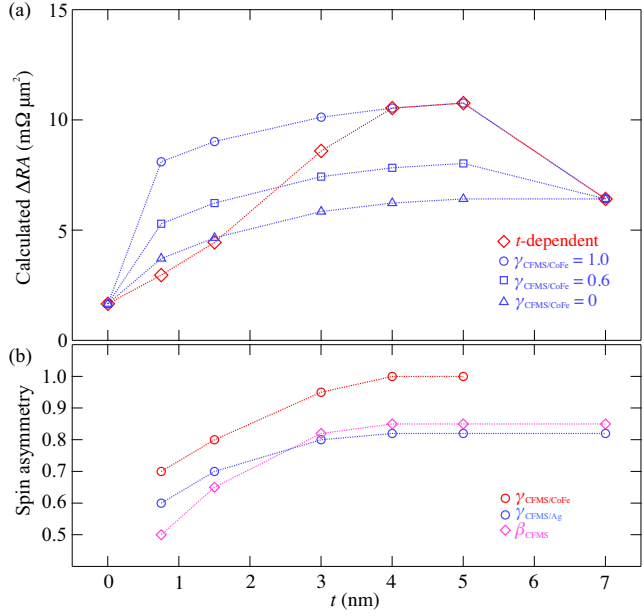}
\caption{(a) Simulated $t$ dependence of $\Delta$$RA$ for Ag(1 nm)/CoFe($7 - t$ nm)/CFMS($t$ nm)/Ag(5 nm)/CFMS($t$ nm)/CoFe($7 - t$ nm)/Ag(5 nm) ($t =$ 0, 0.75, 1.5, 3, 4, 5, and 7 nm) assuming $t$-dependent $\gamma_{\rm CFMS/CoFe}$, $\gamma_{\rm CFMS/Ag}$, and $\beta_{\rm CFMS}$ shown in (b) (red) and $\gamma_{\rm CFMS/CoFe} =$ 0, 0.6, and 1 (blue). (b) Assumed $t$ dependence of $\gamma_{\rm CFMS/CoFe}$, $\gamma_{\rm CFMS/Ag}$, and $\beta_{\rm CFMS}$. All dotted lines are guides to the eye.}
\end{center}
\end{figure}

We attempted to simulate $\Delta$$RA$ as a function of $t$ based on the generalized 2CSR model for metallic multilayers considering the contribution of $\gamma_{\rm CFMS/CoFe}$ \cite{Strelkov_JAP}. 
The 2CSR model assumes individual series resistors for majority- and minority-spin channel, respectively, in metallic multilayers with ferromagnetic layers. 
In this model, resistance of bulk of layers and interfaces and spin diffusion length determine the resistance change of the metallic multilayers when the configuration of magnetization direction of the ferromagnetic layers changes.
The CPP-GMR PSVs for the calculation are set as Ag(1 nm)/CoFe($7 - t$ nm)/CFMS($t$ nm)/Ag(5 nm)/CFMS($t$ nm)/CoFe($7 - t$ nm)/Ag(5 nm) ($t =$ 0, 0.75, 1.5, 3, 4, 5, and 7 nm), where the thickness of the 1-nm-thick bottom Ag layer corresponds to the etched depth of the Ag buffer layer.
For the simulation, it is necessary to input values of the spin-scattering asymmetry coefficients related to the CFMS layer ($\gamma_{\rm CFMS/CoFe}$, $\gamma_{\rm CFMS/Ag}$, $\beta_{\rm CFMS}$), and spin diffusion length of CFMS ($l_{\rm sf}^{\rm CFMS}$).
For $\gamma_{\rm CFMS/Ag}$, $\beta_{\rm CFMS}$, and $l_{\rm sf}^{\rm CFMS}$, we determined $\gamma_{\rm CFMS/Ag} =$ 0.82, $\beta_{\rm CFMS} =$ 0.85, and $l_{\rm sf}^{\rm CFMS} =$ 1.5 nm with reference to the typical or largest analytical values in the previous reports on the CPP-GMR devices with the Co-based Heusler alloys \cite{Nakatani_APL, Sakuraba_PRB, Li_APL}.
Other parameters required for the simulation such as $\gamma$ at the CoFe/Ag ($\gamma_{\rm CoFe/Ag}$); $\beta$ of CoFe layers ($\beta_{\rm CoFe}$); resistivity of Ag ($\rho_{\rm Ag}$), CoFe ($\rho_{\rm CoFe}$), and CFMS ($\rho_{\rm CFMS}$); and spin diffusion length of Ag ($l_{\rm sf}^{\rm Ag}$) and CoFe ($l_{\rm sf}^{\rm CoFe}$) were selected from values in the previous reports: $\gamma_{\rm CoFe/Ag} =$ 0.8 \cite{Jung_JAP}, $\beta_{\rm CoFe} =$ 0.62 \cite{Jung_JAP}, $\rho_{\rm Ag} =$ 2.1 $\mu$$\Omega$ cm \cite{Jung_JAP}, $\rho_{\rm CoFe} =$ 19.1 $\mu$$\Omega$ cm \cite{Delille_JAP}, $\rho_{\rm CFMS} =$ 46 $\mu$$\Omega$ cm \cite{Sakuraba_APL}, $l_{\rm sf}^{\rm Ag} =$ 40 nm \cite{Jung_JAP}, and $l_{\rm sf}^{\rm CoFe} =$ 15 nm \cite{Delille_JAP}.
Further, the $RA$ values for the interfaces were estimated from $\frac{1 + \gamma}{2} \times R^{\uparrow}A$: interfacial $R^{\uparrow}A$ values were determined by considering $R^{\uparrow}A$ for CoFe/CFMS/CoFe, Ag/CFMS/Ag, and Ag/CoFe/Ag estimated from the first-principles calculations as twice interfacial $R^{\uparrow}A$.
We adopt the averaged $R^{\uparrow}A$ for CoFe/CFMS/CoFe and Ag/CFMS/Ag for the possible combinations of terminated atoms at interfaces evaluated using the HAADF-STEM image.

We examined the impact of $\gamma_{\rm CFMS/CoFe}$ on the $t$ dependence of $\Delta$$RA$.
The calculated $\Delta$$RA$ as functions of $t$ assuming $\gamma_{\rm CFMS/CoFe} =$ 0, 0.6, and 1 are shown in Fig. 4(a); $\Delta$$RA$ is enhanced by increasing $\gamma_{\rm CFMS/CoFe}$ and the peak in the $t$ dependence is not present for $\gamma_{\rm CFMS/CoFe} =$ 0, which implies that a finite $\gamma_{\rm CFMS/CoFe}$ needs to be incorporated for reproducing the experimental result in Fig. 3(b).
When $\gamma_{\rm CFMS/CoFe} =$ 1, the simulated $t$ dependence is quantitatively the closest to the experimental result, which implies the presence of a large $\gamma_{\rm CFMS/CoFe}$ at RT. 
Although the peak of $\Delta$$RA$ is reproduced by the simulation, the $t$-dependent behavior of $\Delta$$RA$ in $t <$ 4, in which $\Delta$$RA$ increased significantly by introducing ultrathin ($t =$ 0.75 nm) CFMS layers, seems qualitatively even to be inconsistent with the experimental result. 
Thus, we additionally presumed $t$-dependent $\gamma_{\rm CFMS/CoFe}$, $\gamma_{\rm CFMS/Ag}$, and $\beta_{\rm CFMS}$ as shown in Fig. 4(b), where they were increased and saturated with increasing $t$, taking the degradation of the half metallicity of the Co-based Heusler-alloy film in the range of small thickness into account \cite{Nakatani_Scripta} .
The degradation of the half metallicity of CFMS layers because of the decrease in their thickness was corroborated by the reduction of the inner magnetic moments of Mn atoms, which were confirmed based on the x-ray magnetic circular dichroism \cite{Amemiya_APL, Amemiya_PRB, Tsunegi_PRB} spectra shown in Supplemental Fig. S4 in the supplemental material \cite{Supplemental}. 
The $\Delta$$RA$ was simulated to be increased moderately with increasing $t$ in $t <$ 4 by presuming $t$-dependent $\gamma_{\rm CFMS/Ag}$, $\beta_{\rm CFMS}$, and $\gamma_{\rm CFMS/CoFe}$, as shown by the red diamonds in Fig. 4(a). 
This indicates a more accurate reproduction of the experimental results.
We additionally attempted to calculate $\Delta$$RA$ for $t = 4$ nm and for $\gamma_{\rm CFMS/CoFe} = 1$ under the assumption of larger $\gamma_{\rm CFMS/Ag}$ and $\beta_{\rm CFMS}$ ($\gamma_{\rm CFMS/Ag} = 0.9$ and $\beta_{\rm CFMS} = 0.93$) than the values in the above assumption ($\gamma_{\rm CFMS/Ag} = 0.82$ and $\beta_{\rm CFMS} = 0.85$). As a result, $\Delta$$RA$ is estimated to be 18.1 m$\Omega$ $\mu$m$^{2}$, which is rather quantitatively consistent with the experimentally obtained value ($\Delta$$RA =$ 19.1 m$\Omega$ $\mu$m$^{2}$). Thus, even if the actual values of $\gamma_{\rm CFMS/Ag}$ and $\beta_{\rm CFMS}$ are larger than assumed values that are determined with reference to previous studies, $\gamma_{\rm CFMS/CoFe}$ should still be large and have a large impact on the GMR effect.
Although further experiments and theories are required for more precise simulations, the $t$ dependence of $\Delta$$RA$ in Fig. 3(b) was qualitatively reproduced by the simulation by considering the presence of $\gamma_{\rm CFMS/CoFe}$, the $t$ dependence of the half metallicity of CFMS layers, and the combinations of terminated atoms at the interfaces.
Therefore, we conclude that the $t$-dependent behavior of $\Delta$$RA$ in Fig. 3 is direct evidence for the impact of $\gamma_{\rm CFMS/CoFe}$ on the GMR effect.
It is important to obtain more accurate parameters, particularly $l_{\rm sf}^{\rm CFMS}$, which should be obtained by performing more detailed measurements of $\Delta$$RA$, for quantitatively estimating $\gamma_{\rm CFMS/CoFe}$,. 

In conclusion, we showed the presence of spin-scattering asymmetry at the half-metallic CFMS/CoFe interface, that is, the HMF/FM interface.
The first-principles ballistic transport calculations for (001)-CoFe/CFMS/CoFe showed a large difference of conductance between majority and minority-spin channels, thereby implying a large $\gamma_{\rm CFMS/CoFe}$.
The $\Delta$$RA$ observed in the fully epitaxial CPP-GMR PSVs with top and bottom CFMS($t$ nm)/CoFe($7 - t$ nm) ($0\leq t \leq 7$) layers exhibited $t$ dependence with a clear peak, and it was reproduced qualitatively by the simulation based on the generalized 2CSR model by considering the presence of $\gamma_{\rm CFMS/CoFe}$, the $t$-dependent half metallicity of CFMS layers, and the combinations of terminated atoms at the interfaces confirmed by HAADF-STEM.
We presented direct evidence for the impact of $\gamma_{\rm CFMS/CoFe}$ on the GMR effect by observing the enhancement of $\Delta$$RA$ at RT, which indicates that the HMF/FM interface is expected to yield a large $\gamma$ even at RT.
The introduction of the additional $\gamma$ at the HMF/FM interface by forming an HMF/FM/NM structure leads to further improvement of the MR output in CPP-GMR devices using HMFs. This may contribute to the development of CPP-GMR-based spintronic devices such as CPP-GMR sensors \cite{Nakatani_ieee}

%\section{Acknowledgements}
This work was performed with the approval of the Photon Factory Program Advisory Committee (Proposal No. 2019S2-003). The authors thank B. Dieny for providing the program for the simulation, T. Taniguchi for the fruitful discussion and Prof. K. Amemiya for providing the opportunity to measure XMCD at the KEK Photon Factory. This work was partly supported by Grants-in-Aid for Scientific Research (Nos. 17H06152 and 20H02190) and for Research Activity Start-up (No. 20K22487) from the Japan Society for the Promotion of Science (JSPS).

\makeatletter
\def\maketitle{ %
\say\@authors
\@author@finish
\title@column\titleblock@produce
\suppressfloats[t]%
% Now save some memory.
\let\and\relax
\let\affiliation\@gobble@opt@one
\let\address\affiliation
\let\author\@gobble
\@author@init
\let\@authors\@empty
\let\@authors@curr\@empty
\let\@affil@list\@empty
\let\keywords\@gobble
\let\@keywords\@empty
\let\email\@gobble
\let\@address\@empty
\let\maketitle\relax
\let\thanks\@gobble
\titlepage@sw{ %
\clearpage
 }{}%
 }%
 \makeatother

%\begin{document}
\renewcommand{\figurename}{FIG. S}
\renewcommand{\tablename}{TABLE. S}
\setcounter{figure}{0}
%\preprint{APS/123-QED}
%\begin{center}
%{\LARGE\bf Supplementary Information}
%\end{center}
\title{{Supplemental Material:}  \\ \ Spin-scattering asymmetry at half-metallic ferromagnet/ferromagnet interface}% Force line breaks with \\

\author{Y. Fujita,$^{1}$$\footnote{E-mail: FUJITA.Yuichi@nims.go.jp}$  Y. Miura,$^{2}$ T. Sasaki,$^{2}$ T. Nakatani,$^{2}$ K. Hono,$^{2}$ and Y. Sakuraba$^{2}$$\footnote{E-mail: SAKURABA.Yuya@nims.go.jp}$}
\affiliation{
$^{1}$International Center for Young Scientists, National Institute for Materials Science, 1-2-1 Sengen, Tsukuba, Ibaraki 305-0047, Japan. 
}
\affiliation{
$^{2}$Research Center for Magnetic and Spintronic Materials, National Institute for Materials Science, 1-2-1 Sengen, Tsukuba, Ibaraki 305-0047, Japan. 
}

\date{\today}% It is always \today, today,

%\begin{abstract}
%
%\end{abstract}


\maketitle

%\clearpage
\noindent{\bf FERMI SURFACES OF BULK MATERIALS}

Figures S1 shows the Fermi surfaces in the Brillouin zones of $L2_{1}$-CFMS, bcc-CoFe, and fcc-Ag with the tetragonal unit cell. An overlapping area of the Fermi surface on the $\bm k_{\parallel} = (k_{x}, k_{y})$ between CFMS and CoFe seems to be larger than that between CFMS and Ag due to the absence of the energy surface near $\bm k_{\parallel} = (0, 0)$ of Ag. This results in lower majority-spin conductance at CFMS/CoFe interface compared to that at CFMS/Ag interface.

\begin{figure}[h]
\begin{center}
\includegraphics[width=\textwidth]{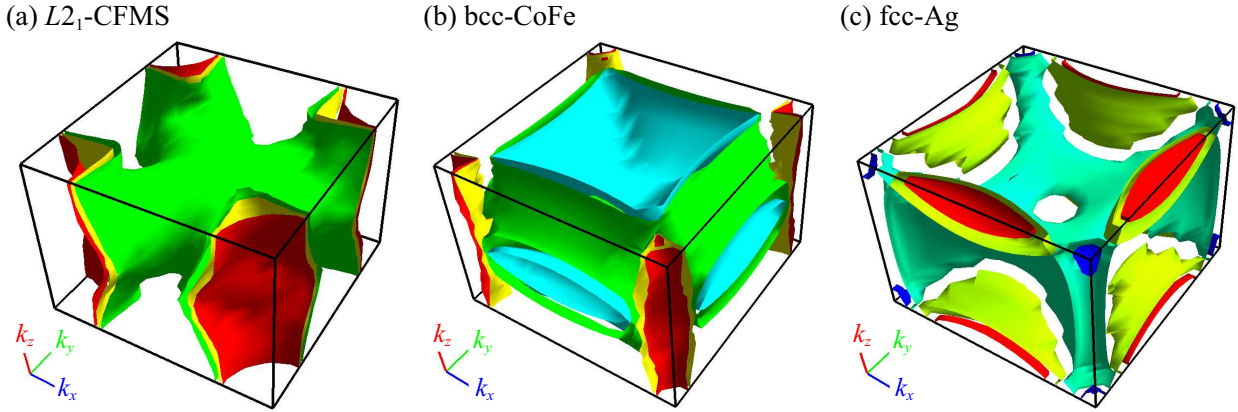}
\caption{Fermi surfaces in the Brillouin zones corresponding to the tetragonal unit cell of (a) $L2_{1}$-CFMS, (b) bcc-CoFe, and (c) fcc-Ag plotted by FermiSurfer \cite{FermiSurfer}. }
\end{center}
\end{figure}

\noindent{\bf DETAILED GROWTH PROCEDURE OF CPP-GMR PSV}

We formed fully epitaxial CoFe/CFMS/Ag/CFMS/CoFe structured CPP-GMR PSVs with the whole structure of MgO(001) single-crystalline substrate/Cr (20 nm)/Ag (80 nm)/Co$_{50}$Fe$_{50}$ ($7 - t$ nm)/CFMS ($t$ nm)/Ag (5 nm)/CFMS ($t$ nm)/Co$_{50}$Fe$_{50}$ ($7 - t$ nm)/Ag (5 nm)/Ru (8 nm) with various $t$ ($t =$ 0, 0.75, 1.5, 3, 4, 5, and 7 nm).
All layers were deposited using DC or RF sputtering methods at room temperature (RT) and in an ultrahigh vacuum sputtering system comprising two connected chambers for DC and RF sputtering with base pressures of $\sim$$8 \times 10^{-7}$ Pa and $\sim$$3 \times 10^{-6}$ Pa, respectively. 
Sputtering type (DC or RF), sputtering pressure, and deposition rate are summarized in Table SI.

\begin{table}[h]
	\begin{center}
		\caption{Sputtering type, pressure, and rate for each layer.}
		%\label{table: 2-2_samples}
		%\small
		\vspace{2mm}
  		\begin{tabular}{c ccc}
			\hline
			\hline
			Layer &\ \ Type &\ \ Pressure (Pa) &\ \ Rate (\AA/s)\\
			\hline
			Cr & DC & $\sim$0.09 & 0.18\\
			Ag & RF & $\sim$0.10 & 0.40\\
			CoFe & DC & $\sim$0.20 & 0.25\\
			CFMS & DC & $\sim$0.11 & 1.15\\
			Ru & RF & $\sim$0.18 & 0.41\\
			\hline
			\hline
		\end{tabular}
	\end{center}
\end{table}

The CoFe, CFMS, and Cr layers were deposited by DC sputtering, and Ag and Ru layers were deposited by RF sputtering. 
Post-deposition annealing was performed $in$ $situ$.
An MgO(001) single-crystalline substrate cleaned using acetone, isopropanol, and diluted water in an ultrasonicator in advance was annealed at 600 $^{\circ}$C for 30 min and cooled for 2 h. 
Next, Cr (20 nm)/Ag (80 nm) buffer layers were deposited on the MgO(001) single-crystalline substrate. 
After deposition, the sample was annealed at 300 $^{\circ}$C for 30 min to suppress surface roughness. 
Then, a bottom CoFe ($7 - t$ nm) layer was deposited, followed by the deposition of a CFMS ($t$ nm) layer. 
The CFMS layer was deposited using a co-sputtering method with targets of Co$_{42.8}$Mn$_{29.4}$Si$_{27.9}$ alloy, Co$_{47.5}$Fe$_{24.2}$Si$_{28.4}$ alloy, and Si. 
Using X-ray florescence, the actual chemical composition of a CFMS film formed under the same condition was confirmed to be Co$_{48.9}$Fe$_{14.9}$Mn$_{9.4}$Si$_{26.8}$, which indicates the formation of a stoichiometric CFMS layer. 
After the deposition of the CFMS layer, the sample was annealed at 450 $^{\circ}$C for 2 min to promote atomic ordering. 
After cooling, a 5-nm-thick Ag spacer layer was deposited on the CFMS layer (or on the CoFe layer for $t$ = 0 nm); then, the top CFMS ($t$ nm)/CoFe ($7 - t$ nm) layers were formed under the same condition as the bottom CFMS/CoFe layers. 
After the deposition of the top CFMS layer, the sample was annealed at 450 $^{\circ}$C for 2 min. 
Finally, after cooling, a Ag (5 nm) layer and a Ru (8 nm) capping layer were deposited on top of the CoFe layer (or on the CFMS layer for $t$ = 7 nm). 

%\vspace{5mm}

%\subsection{}
\noindent{\bf STRUCTURAL ANALYSIS OF CPP-GMR PSV}

\begin{figure}[b]
\begin{center}
\includegraphics[width=13cm]{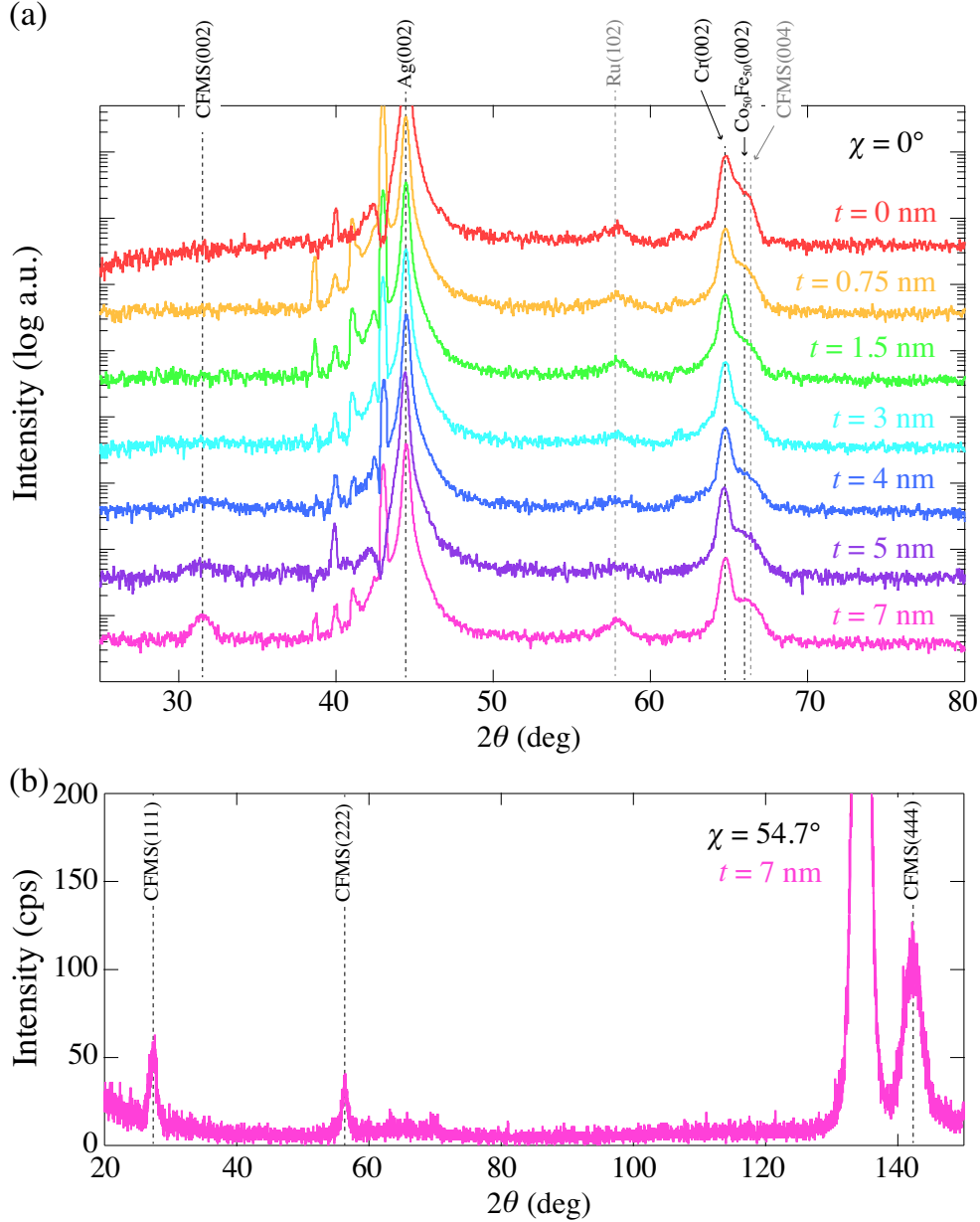}
\caption{$\theta$-$2\theta$ XRD patterns for the formed CPP-GMR PSVs obtained by setting the scattering vector to the (a) out-of-plane ($\chi =$ 0$^{\circ}$) direction for $t =$ 0, 0.75, 1.5, 3, 4, 5, and 7 nm and (b) (111) plane ($\chi =$ 54.7$^{\circ}$) for $t =$ 7 nm.}
\end{center}
\end{figure}

Figure. S2(a) shows the X-ray diffraction (XRD) patterns of $\theta$-$2\theta$ scans from the $\langle$001$\rangle$ directions ($\chi =$ 0$^{\circ}$) of the samples with $t =$ 0, 0.75, 1.5, 3, 4, 5, and 7 nm.
We found diffraction peaks from the (001) plane of Ag, Cr, and CoFe for all $t$. Further, we confirmed a (002) superlattice peak from the $B2$ ordering of the CFMS layers for $t =$ 4, 5, and 7 nm.
For $t =$ 7 nm, we observed (111) superlattice peaks originating from the $L2_{1}$ ordering of CFMS layers in a $\theta$-$2\theta$ XRD profile captured from $\langle$111$\rangle$ directions set by tilting the sample to $\chi =$ 54.7$^{\circ}$, as shown in Fig. S2(b).
Although the intensities were too weak to detect either (002) or (111) peaks for $t <$ 3 and $t <$ 7, respectively, the $L2_{1}$ phases in the CFMS layers were confirmed even under regimes where they were thin via detailed structural analyses using a high-angle annular dark-field scanning transmission electron microscope (HAADF-STEM) and nanobeam electron diffraction (NED).
We successfully formed the fully epitaxial CPP-GMR PSVs with $L2_{1}$-ordered CFMS layers.

\begin{figure}[t]
\begin{center}
\includegraphics[width=\textwidth]{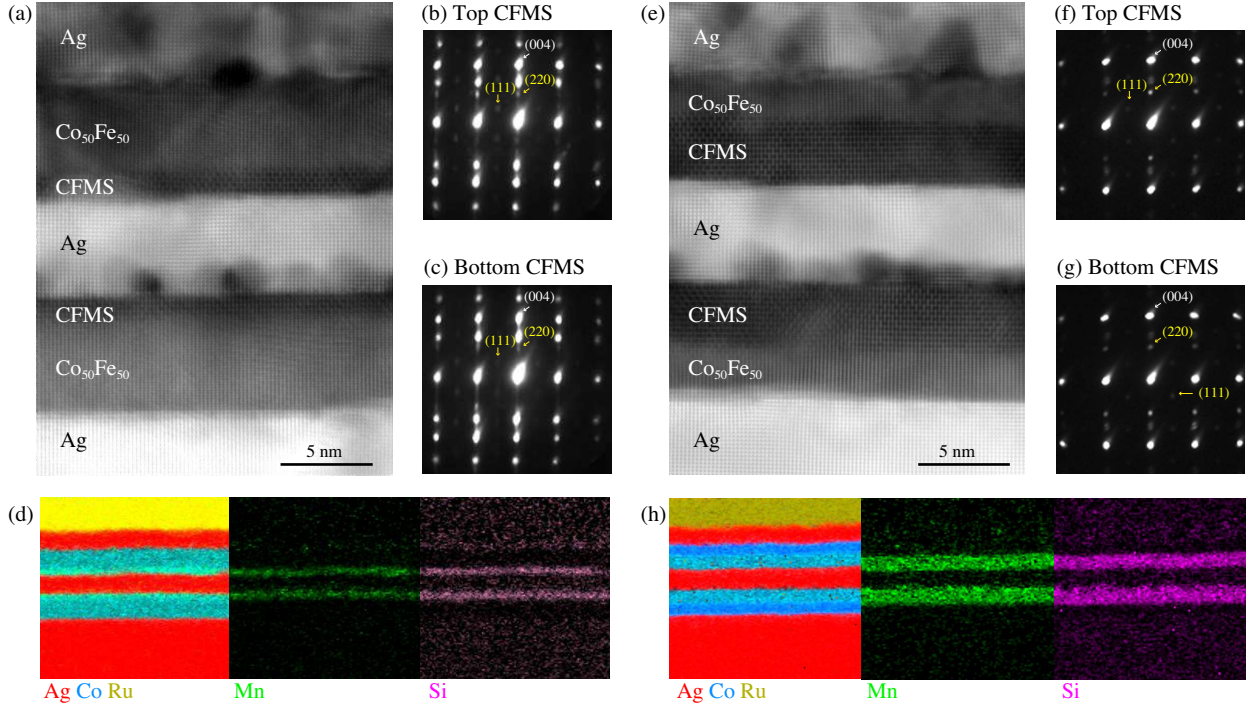}
\caption{Cross-sectional HAADF-STEM images for (a) $t =$ 1.5 nm and (e) $t =$ 4 nm, NED patterns for top and bottom CFMS layers for (b), (c) $t =$ 1.5 nm and (f), (g) $t =$ 4 nm, and EDS elemental maps for (d) $t =$ 1.5 nm and (h) $t =$ 4 nm, respectively.}
\end{center}
\end{figure}

Figures S3(a) and S3(e) show cross-sectional HAADF-STEM images for the PSVs with $t =$ 1.5 nm and 4 nm, respectively. 
All layers and interfaces are flat and smooth in both samples, and we can distinctly recognize the CFMS/CoFe interfaces.
In the NED patterns taken from the [011] zone axis of the top and bottom CFMS layers for $t = $ 1.5 nm [Figs. S3(b) and S3(c)] and $t = $ 4 nm [Figs. S3(f) and S3(g)], $\{$001$\}$ and $\{$111$\}$ superlattice spots corresponding to the $L2_{1}$ phase are observed.
Thus, the CFMS layers in the samples for both $t =$ 1.5 nm and 4 nm are expected to have the $L2_{1}$-ordered phase.
Figures S3(d) and S3(h) show the energy dispersive X-ray spectroscopy (EDS) elemental maps of Ru, Co, Ag, Mn, and Si in the CPP-GMR PSVs with $t =$ 1.5 nm and $t =$ 4 nm, respectively.
These elemental maps indicate that no significant atomic diffusion occurs in both samples. 

\vspace{5mm}

\noindent{\bf ANALYSIS OF MR OUTPUT}

\begin{figure}[b]
\begin{center}
\includegraphics[width=15cm]{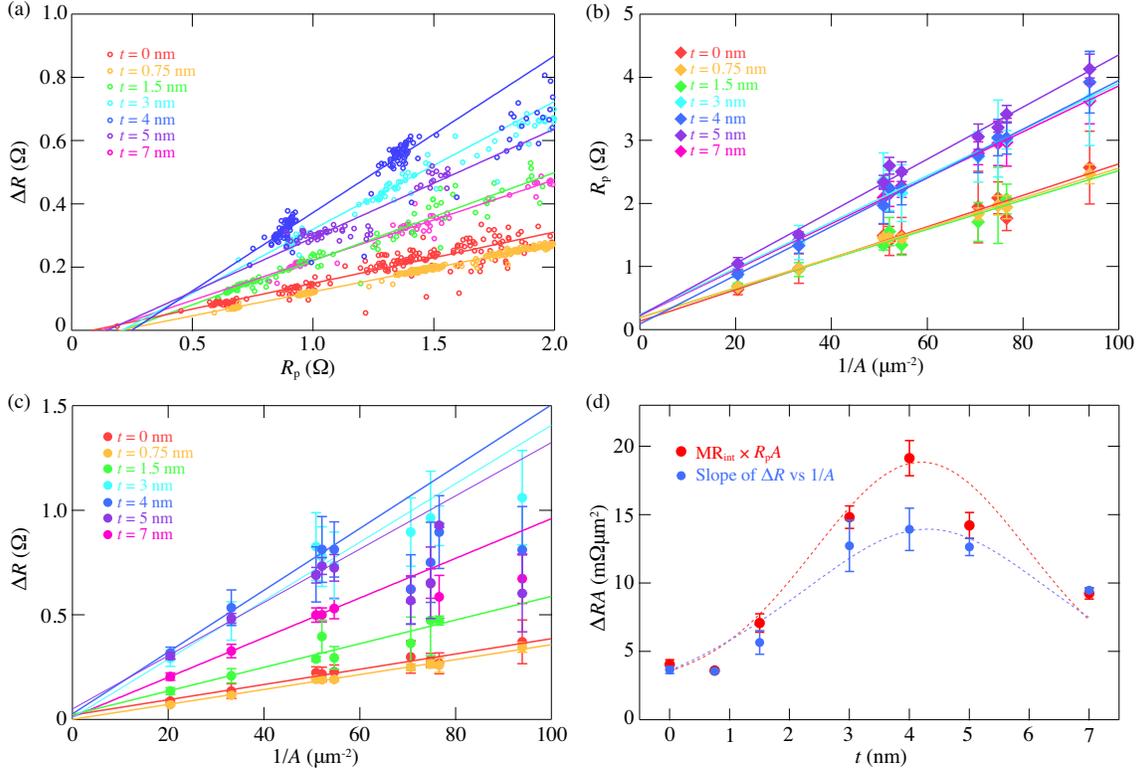}
\caption{(a) $\Delta$$R$ vs. $R_{\rm p}$ plots, (b) $R_{\rm p}$ vs. $1/A$ plots, (c) $\Delta$$R$ vs. $1/A$ plots, and (d) $t$ dependence of $\Delta$$RA$.}
\end{center}
\end{figure}

We estimated MR$_{\rm int}$ and $R_{\rm p}A$ from $\Delta$$R$ vs. $R_{\rm p}$ plots [Fig. S4(a)] and $R_{\rm p}$ vs. $1/A$ plots [Fig. S4(b)], respectively.
$R_{\rm p}$ and $\Delta R = R_{\rm ap} - R_{\rm p}$ can be read from the observed MR curves.
$A$ represents the averaged actual size of some representative pillars for each designed pillar size. Since the pillar size was roughly deduced by analyzing the scanning electron microscope images, the used $A$ had an analysis error.
In addition, in the plots with the $1/A$ axis, we used the averaged $\Delta R$ and $R_{\rm p}$ values corresponding to each averaged $A$.
From the slopes of $\Delta$$R$ vs. $1/A$ plots [Fig. S4(c)], we obtained $\Delta RA$ as a function of $t$, as indicated by the blue circles in Fig. S4(d).
The quantitative difference in $\Delta$$RA$ between that estimated from MR$_{\rm int}$ $\times$ $R_{\rm p}A$ and from $\Delta$$R$ vs. $1/A$ plots is shown in Fig. S4(d).
MR$_{\rm int}$ $\times$ $R_{\rm p}A$ can be expressed as $\frac{\Delta R^{\rm m}}{R_{\rm p}^{\rm m} - R_{\rm para}}$ $\times$ $\frac{R_{\rm p}^{\rm m} - R_{\rm para}^{\prime}}{1/A}$, where $R_{\rm para}$ and $R_{\rm para}^{\prime}$ are parasitic resistances that correspond to the $X$-intercepts of $\Delta$$R$ vs. $R_{\rm p}$ plots and the $Y$-intercepts of $R_{\rm p}$ vs. $1/A$ plots, respectively.
$\Delta R^{\rm m}$ and $R_{\rm p}^{\rm m}$ were read values of $\Delta R$ and $R_{\rm p}$ from the observed MR curves, respectively. $\Delta$$RA$ estimated from the slopes of $\Delta$$R$ vs. $1/A$ plots can be expressed as $(\Delta R^{\rm m} - \Delta R^{\prime})A$, where $\Delta R^{\prime}$ corresponds to the finite $Y$-intercepts of $\Delta R$ vs. $1/A$ plots unwillingly yielded in the analyses.
MR$_{\rm int}$ $\times$ $R_{\rm p}A$ should quantitatively be in agreement with $\Delta RA$ obtained from the $\Delta$$R$ vs. $1/A$ plots when $R_{\rm para} = R_{\rm para}^{\prime}$ and $\Delta R^{\prime} = 0$.
The finite deviation of parasitic resistances and $\Delta R^{\prime}$ expected to cause a quantitative difference in $\Delta$$RA$ associated with the difference in the estimation method were inevitably caused in the analyses. This is because the $A$ values used here had the analysis error and were not individual values for each pillar but the averaged actual size of some representative pillars.

\vspace{5mm}
%\clearpage

\noindent{\bf X-RAY MAGNETIC CIRCULAR DICHROISM}

\begin{figure}[b]
\begin{center}
\includegraphics[width=\textwidth]{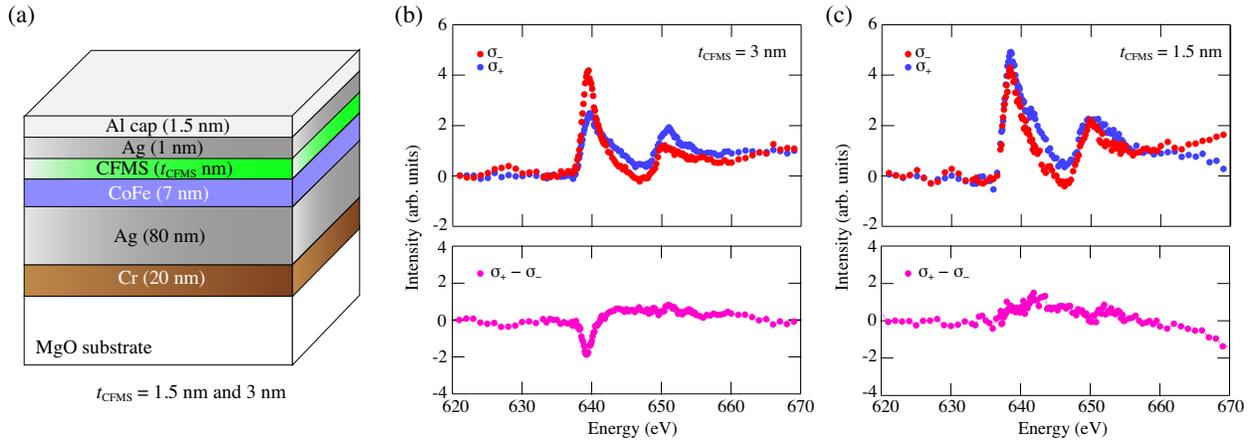}
\caption{(a) Sample structure for XMCD measurements. XA (upper) and XMCD (lower) spectra at Mn $L_{2, 3}$ edges for (b) $t_{\rm CFMS} =$ 3 nm and (c) $t_{\rm CFMS} =$ 1.5 nm measured at RT.}
\end{center}
\end{figure}

We performed measurements of X-ray magnetic circular dichroism (XMCD) to verify the magnetic properties of CFMS layers. 
We prepared fully epitaxial samples for XMCD measurement comprising MgO(001) single-crystalline substrate/Cr (20 nm)/Ag (80 nm)/CoFe (7 nm)/CFMS ($t_{\rm CFMS}$ nm)/Ag (1 nm)/Al (1.5 nm) with $t_{\rm CFMS} =$ 1.5 and 3 nm as displayed in Fig. S5(a).
We measured X-ray-absorption (XA) and XMCD spectra in total electron yield (TEY) mode at RT in an ultrahigh vacuum chamber at BL 7A at the Photon Factory of High-energy Accelerator Research Organization, Japan \cite{Tsunegi_PRB}. 
The samples were in a saturated magnetization state and oriented along the X-ray propagation direction. 
The Mn $L$-edge XA spectra were measured with the field parallel and antiparallel to the fixed photon helicity, and the XMCD spectra were deduced from the difference between the two spectra.
Figures S5(b) and S5(c) show XA (upper) and XMCD (lower) spectra of the inner layers of CFMS around Mn $L_{2, 3}$-absorption edges in the samples with $t_{\rm CFMS} =$ 3 nm and $t_{\rm CFMS} =$ 1.5 nm, respectively.
The XMCD spectra were surely observed for $t_{\rm CFMS} =$ 3 nm (peak around $\sim$639 eV). For $t_{\rm CFMS} =$ 1.5 nm, we could not see an evident XMCD spectra. 
Thus, the magnetic moment of Mn atoms is expected to be lowered by reducing the thickness of the CFMS layers.
Unfortunately, in this XMCD measurement, background matching between $\sigma_{-}$ and $\sigma_{+}$ spectra does not seem to be good. This may be caused by attenuation of TEY signal owing to the presence of Ag and Al capping layers. 

%\clearpage

%\end{document}
%
% ****** End of file apssamp.tex ******

\end{document}